\documentclass[aip,jap,reprint,groupedaddress,showpacs]{revtex4-1}

\usepackage{graphicx}
\usepackage{dcolumn}
\usepackage{bm}

\begin{document}

\title{Large magnetic heat transport in a Haldane chain material
Ni(C$_3$H$_{10}$N$_2$)$_2$NO$_2$ClO$_4$}

\author{X. F. Sun$^{1}$}
\email{xfsun@ustc.edu.cn}

\author{X. G. Liu$^{1}$}

\author{L. M. Chen$^{2}$}

\author{Z. Y. Zhao$^{1}$}

\author{X. Zhao$^{3}$}

\affiliation{$^{1}$Hefei National Laboratory for Physical Sciences
at Microscale, University of Science and Technology of China,
Hefei, Anhui 230026, People's Republic of China}

\affiliation{$^{2}$Department of Physics, University of Science
and Technology of China, Hefei, Anhui 230026, People's Republic of
China}

\affiliation{$^{3}$School of Physical Sciences, University of
Science and Technology of China, Hefei, Anhui 230026, People's
Republic of China}


\begin{abstract}

We report a study on the heat transport of an $S = 1$ Haldane
chain compound Ni(C$_3$H$_{10}$N$_2$)$_2$NO$_2$ClO$_4$ at low
temperatures and in magnetic fields. The zero-field thermal
conductivities show a remarkable anisotropy for the heat current
along the spin-chain direction ($\kappa_b$) and the vertical
direction ($\kappa_c$), implying a magnetic contribution to the
heat transport along the spin-chain direction. The
magnetic-field-induced change of the spin spectrum has obviously
opposite impacts on $\kappa_b$ and $\kappa_c$. In particular,
$\kappa_b(H)$ and $\kappa_c(H)$ curves show peak-like increases
and dip-like decreases, respectively, at $\sim$ 9 T, which is the
critical field that minimizes the spin gap. These results indicate
a large magnetic thermal transport in this material.

\end{abstract}


\maketitle

The heat transport in one-dimensional (1D) quantum magnets has
attracted much attention due to the role of magnetic
excitations.\cite{Heidrich1, Hess1, Sologubenko1} Theoretical and
experimental results have made an agreement that the $S$ = 1/2
Heisenberg chain system, which is an integrable spin model,
exhibits a ballistic transport of the spin
excitations.\cite{Zotos2, Narozhny, Zotos3, Heidrich2,
Sologubenko3, Hess2} However, the situation of the $S$ = 1 chain
system is not clear yet. Several theories on this nonintegrable
spin model predicted either a diffusive or a ballistic transport
of spin excitations.\cite{Zotos2, Fujimoto, Orignac, Karadamoglou}
The experimental results on the $S$ =1 Haldane chain materials
AgVP$_2$S$_6$ and Y$_2$BaNiO$_5$ showed considerably small
magnetic thermal conductivity and therefore seemed to support the
model of diffusive spin transport.\cite{Sologubenko4, Kordonis}
However, a large magnetic thermal conductivity has been revealed
in a two-leg Heisenberg $S$ = 1/2 ladder compound
(La,Sr,Ca)$_{14}$Cu$_{24}$O$_{41}$,\cite{Sologubenko5, Hess3}
which potentially shows the evidence of a ballistic behavior. This
is surprising because the $S$ = 1 chain and the $S$ = 1/2 ladder
are essentially the same in the aspects of the spin-liquid ground
state and the gapped magnetic spectrum.\cite{White} In a recent
work on an organic $S$ = 1 Haldane chain compound
Ni(C$_2$H$_8$N$_2$)$_2$NO$_2$ClO$_4$ (abbreviated as NENP), which
has relatively weaker spin interaction and smaller spin gap
($\sim$ 12.2 K),\cite{Lu, Regnault} the magnetic heat transport
was found to be rather large.\cite{Sologubenko6} Since the spin
transport of NENP can only be observed in magnetic field, which
weakens the spin gap,\cite{Sologubenko6} it is possible that the
large spin transport in zero magnetic field can be found in $S$ =
1 chain systems with smaller energy gaps.

Ni(C$_3$H$_{10}$N$_2$)$_2$NO$_2$ClO$_4$ (abbreviated as NINO),
which was found to be another ideal $S$ = 1 Haldane chain system,
has a similar spin structure to that of NENP (see Fig. 1). The
Ni$^{2+}$ spins ($S$ = 1) form the spin chains along the $b$ axis,
in which the intrachain antiferromagnetic (AF) interaction ($J
\approx 50$ K) is about a factor of $10^4$ times stronger than the
interchain interaction ($J'$).\cite{Renard, Takeuchi, Sieling,
Hagiwara} It is known that in an isotropic AF $S$ = 1 chain
system, the spin excitations are triply degenerate with an energy
gap $E_g \approx 0.41 J$, which is about 20.3 K for
NINO.\cite{Sakai} However, due to the strong planar anisotropy and
weak orthorhombic anisotropy, the Haldane gap is split into three
gaps with zero-field values $\Delta E_1 = 8.3$ K, $\Delta E_2 =
12.5$ K and $\Delta E_3 = 21.9$ K.\cite{Takeuchi} When an external
magnetic field is applied along the $a$ axis, $\Delta E_2$ keeps
constant, $\Delta E_3$ increases, and $\Delta E_1$ decreases. The
smallest gap $\Delta E_1$ is apparently the most important for the
low-energy magnetic excitations. In particular, $\Delta E_1$
descends to a small value at a critical field $H_c \approx$ 9 T,
and then increases above $H_c$.\cite{Sieling, Hagiwara} In
general, magnetic properties of NINO are very similar to those of
NENP, except that the energy scales of the spin gap are different.

In this work, we study the heat transport of NINO single crystals
at low temperatures down to 0.3 K and in magnetic fields up to 14
T. It is found that thermal conductivities show a remarkable
anisotropy between the direction along the spin chain and that
perpendicular to it, which indicates a large magnetic contribution
to the heat transport along the spin-chain direction.

High-quality NINO single crystals were grown by a slow evaporation
method from aqueous solution.\cite{Tao} The largest surfaces of
the as-grown crystals are parallel to the $bc$ plane, confirmed by
the X-ray diffraction. Therefore, it is possible to obtain large
parallelepiped-shaped samples with the longest dimension along the
$b$ axis or the $c$ axis. The thermal conductivities were measured
using a conventional steady-state technique along the $b$ axis
($\kappa_b$) and the $c$ axis ($\kappa_c$), for two samples with
sizes of 3.5 $\times$ 1.68 $\times$ 0.59 {mm}$^3$ and 3.1 $\times$
1.50 $\times$ 0.51 {mm}$^3$, respectively. Details for the
measurements have been described elsewhere.\cite{Sun_DTN,
Zhao_Nd2CuO4, Zhao_BCVO}

\begin{figure}
\includegraphics[clip,width=8.5cm]{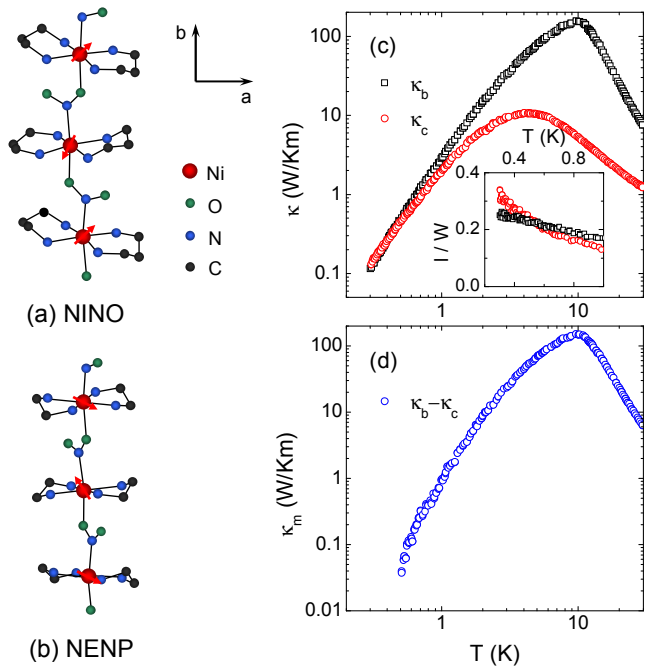}
\caption{(color online) (a,b) Schematic view of the spin-chain
structure of NINO and NENP (Ref. \onlinecite{Renard}). In zero
field, the Ni$^{2+}$ spins show a (disordered) ground state of the
Haldane gap state. (c) Temperature dependencies of the thermal
conductivities $\kappa_b$ and $\kappa_c$ of NINO single crystals
in zero magnetic field. The inset shows the temperature
dependencies of the ratio of the phonon mean free path $l$ to the
averaged sample width $W$. (d) Magnetic thermal conductivity
$\kappa_m(T)$ obtained from $\kappa_b - \kappa_c$.}
\end{figure}

Figure 1(c) shows the temperature dependencies of $\kappa_b$ and
$\kappa_c$ of NINO single crystals in zero magnetic field.
Apparently, the behaviors of $\kappa_b$ and $\kappa_c$ seem to be
different from usual phonon transport properties of
insulators.\cite{Berman} In nonmagnetic insulators, the position
and the magnitude of the low-$T$ phonon peak are mainly determined
by the impurity/defect scattering on phonons. Therefore, the
phonon peaks are usually located at similar temperatures for
different directions of heat currents and the shapes (temperature
dependencies) of peaks are almost identical for different
directions. These are clearly different to what the NINO data
show. The most remarkable phenomenon in Fig. 1(c) is that the
$\kappa_b$ and $\kappa_c$ differ significantly at relatively high
temperatures, but they are almost the same at subKelvin
temperatures. All these features suggest that the heat transport
of NINO cannot be a simple phononic behavior. One may naively
expect that the difference is due to the additional magnetic
thermal transport along the chain direction. However, the
separation of  phononic and magnetic conductivities from the
experimental data is not so straightforward. A common way to
obtain the magnetic thermal conductivity in the quasi-1D magnets
is to assume that the phonon term along the spin-chain direction
can be estimated from the thermal conductivity perpendicular to
the spin chain.\cite{Hess1, Sologubenko1} The main uncertainty of
this method is related to the anisotropy of the phonon thermal
transport itself, which is known to be insignificant for most
materials (with some exceptions like those having layered
lattices).\cite{Berman}

In an earlier work, another Haldane chain compound, NENP, was
found to exhibit an anisotropic phonon thermal conductivity even
at subKevin temperatures.\cite{Sologubenko6} The magnetic thermal
conductivities of NENP were obtained from the
magnetic-field-induced increase of $\kappa$ along the spin
chain.\cite{Sologubenko6} In our data, however, the $\kappa_b$ and
$\kappa_c$ are nearly the same at very low temperatures. This
indicates an isotropic phonon transport at such low temperatures,
where the magnetic excitations are negligible. In this regard, it
is useful to estimate the mean free path of phonons, $l$, by using
the kinetic formula $\kappa_{ph} = \frac{1}{3}Cv_pl$, where $C =
\beta T^3$ is the phonon specific heat at low temperatures and
$v_p$ is the averaged sound velocity.\cite{Zhao_Nd2CuO4, Comment}
The phonon specific-heat coefficient $\beta$ is known from the
specific heat measurements,\cite{Tao} and $v_p$ can be obtained
from $\beta$. The inset to Fig. 1(c) shows the temperature
dependencies of the ratio $l/W$ for $\kappa_b$ and $\kappa_c$,
where $W$ is the averaged sample width. It is found that in both
cases the $l$ becomes comparable to $W$ at $T \rightarrow$ 0.3 K,
indicating that the phonon heat transport approaches the boundary
scattering limit.\cite{Berman, Zhao_Nd2CuO4, Comment} With
increasing temperature, the $\kappa_b$ and $\kappa_c$ show a large
difference because the magnetic excitations become populated and
they can either transport heat along the spin-chain direction or
scatter phonons in other directions.

We can estimate the magnetic thermal conductivity along the spin
chain by subtracting the $\kappa_c$ from $\kappa_b$. As shown in
Fig. 1(d), the magnetic thermal conductivity is very large and
reaches a value of $\sim$ 150 W/Km at 10 K. Thus, this compound
has larger thermal conductivity than many other low-dimensional
quantum magnets.\cite{Heidrich1, Hess1, Sologubenko1,
Sologubenko4, Kordonis, Sologubenko6}

\begin{figure}
\includegraphics[clip,width=6.0cm]{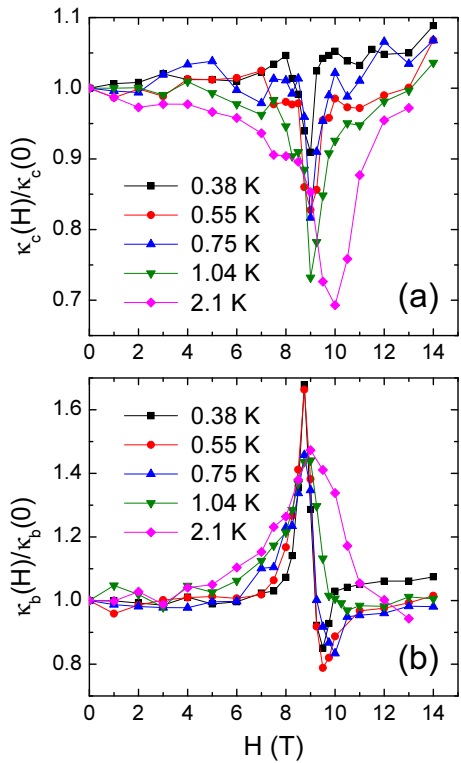}
\caption{(color online) Magnetic-field dependencies of $\kappa_c$
and $\kappa_b$ of NINO single crystals in magnetic field parallel
to the $a$ axis.}
\end{figure}

The magnetic-field dependencies of $\kappa_b$ and $\kappa_c$ have
been measured at low temperatures and in fields along the $a$
axis. As shown in Fig. 2, both the $\kappa_b$ and $\kappa_c$ show
weak field dependencies except in the vicinity of 9 T, which is
known to be the critical field of this material. Apparently, at
the critical field, the spin gap is minimized and therefore the
magnetic excitations are well populated. The dip-like
$\kappa_c(H)$ clearly shows the strong suppression of phonon
transport due to scattering by magnetic excitations. This result
indicates that magnetic excitations also play a role in the heat
transport as phonon scatterers in this direction. In the
spin-chain direction, the increase of magnetic excitations leads
to a peak of $\kappa_b(H)$ at the critical field. This again
demonstrates that the magnetic excitations in this material can
transport heat directly in the spin-chain direction. Therefore,
the $\kappa(H)$ behaviors shown in Fig. 2 are supportive for the
above understanding of the zero-field heat transport data. At very
low temperatures, the $\sim$ 8 K gap impedes the low-energy
magnetic excitations and the phonon transport is dominated, which
results in a nearly isotropic thermal transport. With increasing
temperature, the magnetic excitations become more populated and
they take part in the heat transport by acting as heat carriers in
the spin-chain direction and as phonon scatterers in other
directions. Thus, the $\kappa_b$ and $\kappa_c$ show a large
difference at relatively high temperatures. The present
experimental phenomena are very similar to those in NENP. However,
the effect of magnetic excitations scattering phonons was not
observed in NENP.\cite{Sologubenko6} Another difference is that
the relative increase of $\kappa$ at $H_c$ is 4--6 times larger in
NENP. However, since the absolute magnitude of $\kappa$ is more
than 10 times smaller in NENP, the field-induced increase of
magnetic thermal conductivity is actually stronger in NINO.

In summary, we observe a strongly anisotropic heat transport along
the spin-chain direction and the vertical direction in NINO single
crystals. This indicates that the $S$ = 1 chain system can also
have larger magnetic thermal conductivity. It is essentially
consistent with the recent results on another Haldane chain
compound, NENP, in which magnetic thermal transport was evidenced
when the spin gap was suppressed by magnetic
field.\cite{Sologubenko6} It is necessary to investigate the
origin of the remarkably different ability of spin transport in
many $S$ = 1 chain compounds.

This work was supported by the National Natural Science Foundation
of China, the National Basic Research Program of China (Grant Nos.
2009CB929502 and 2011CBA00111), and the Fundamental Research Funds
for the Central Universities (Program No. WK2340000035).

\end{document}